# UNIFYING DESIGN-BASED AND MODEL-BASED SAMPLING THEORY — SOME SUGGESTIONS TO CLEAR THE COBWEBS

BEN O'NEILL[*], *ACIL Allen*[**]

WRITTEN 24 AUGUST 2024

**Abstract**

This paper gives a holistic overview of both the design-based and model-based paradigms for sampling theory. Both methods are presented within a unified framework with a simple consistent notation, and the differences in the two paradigms are explained within this common framework. We examine the different definitions of the "population variance" within the two paradigms and examine the use of Bessel's correction for a population variance. We critique some messy aspects of the presentation of the design-based paradigm and implore readers to avoid the standard presentation of this framework in favour of a more explicit presentation that includes explicit conditioning in probability statements. We also discuss a number of confusions that arise from the standard presentation of the design-based paradigm and argue that Bessel's correction should be applied to the population variance.

DESIGN-BASED SAMPLING THEORY; MODEL-BASED SAMPLING THEORY; SAMPLE VARIANCE; POPULATION VARIANCE; BESSEL'S CORRECTION; CONDITIONING; PRESENTATION.

The goal of this paper is to give a holistic view of sampling theory that includes both the design-based and model-based paradigms. The author has observed a great deal of confusion among students and novice practitioners in relation to these competing paradigms in sampling theory. This confusion is unsurprising, since these paradigms and the textbooks that use them often use identical notation to refer to different mathematical objects and use identical syntactical presentation of formulae to assert completely different things. Various sampling theory textbooks that adopt these two paradigms assert formulae using identical notation, but with objects defined differently, and this leads to results that are *prima facie* contradictory. This confusion occurs largely for historical reasons — the design-based paradigm is older than the model-based paradigm and it uses notational conventions that are now somewhat antiquated.

In order to explain the two sampling theory paradigms clearly —and address common areas of confusion— this paper sets out the core results of both paradigms in an explicit notation that encompasses both cases within the one unifying framework. Our unifying framework involves explicit use of conditioning statements in probability and moment results, which ensures that the core results in the two paradigms are differentiated and can be related clearly. Part of our analysis and unification involves an examination of the notion of a "population variance". We analyse this concept and provide clarity of what constitutes a sensible definition of this concept.

---

[*] E-mail address: ben.oneill@hotmail.com.
[**] ACIL Allen, Level 6, 54 Marcus Clarke Street, Canberra ACT 2601.

In particular, we look at the application of Bessel's correction in the population variance, which is a common source of confusion and variation in practice in this field. We critique some messy aspects of the presentation of the design-based paradigm and show how it can be presented more clearly and understood more simply.

As part of our analysis, we will advance the view that Bessel's correction **should be applied** to the population variance and that there is no logically consistent basis for the folk practice of excluding this term when defining the population variance. Cochran (1977) notes the variation in convention for defining the population variance, noting that the variance of the sample units in a finite population is usually defined (within the design-based literature) without Bessel's correction, but that the core results in sampling theory are simpler when stated in terms of a corresponding expression using Bessel's correction (p. 23). He states that the inclusion of Bessel's correction "…has been used by those who approach sampling theory by means of the analysis of variance. Its advantage is that most results take a slightly simpler form." Barnett (1974) likewise notes the variation in practice and observes that use of Bessel's correction "…is particularly convenient in that it leads to simpler algebraic expressions … and produces results more closely resembling corresponding ones in the infinite population context." (p. 24)

Our analysis and suggestions will be primarily focused on proper conditioning in probabilistic results and the inclusion or absence of Bessel's correction in the population variance, yielding a clear framing of all relevant results. The title of this paper could just as easily have been "classical design-based sampling theory is a confusing mess with antiquated notation", but we have instead opted to frame the paper in more uplifting terms, seeking to "unify" the two paradigms. Following the classical "design-based" view of sampling theory, many statistical textbooks give bad or confusing advice or bad notation and presentation of results, which can leads to confusion and incorrect practice. The goal of the present paper is to dispel confusion surrounding this issue and make suggestions for future practice.

**1. Descriptive quantities for the sample, population and superpopulation**

This paper will discuss both the model-based and design-based paradigms and look at the issue of Bessel's correction in both contexts. To do this clearly and effectively, we must first adopt a single consistent notation that allows clear reference to all the relevant descriptive objects that arise in both paradigms. To do this we adopt the notation of the model-based paradigm to differentiate objects describing the superpopulation from those describing the finite population. (However, we will make note of the alternative notation used in the design-based method when needed.) We examine the following three-level hierarchy:

$$\text{Superpopulation} \quad \quad \boldsymbol{X} = (X_1, X_2, X_3, \ldots),$$
$$\text{Population} \quad \quad \boldsymbol{X}_N = (X_1, X_2, \ldots, X_N),$$
$$\text{Sample} \quad \quad \boldsymbol{X}_n = (X_1, X_2, \ldots, X_n).$$

Our superpopulation is an infinite sequence of random variables which contains the population and sample. We will consider a population with positive finite size $N$ and a sample of positive size $n$, giving us the restriction $1 \leq n \leq N \leq \infty$.[1] (For most of our analysis we will want to be able to compute the sample variance incorporating Bessel's correction so we will assume that $n \geq 2$ in these cases.)

To describe the overall "model" and results for sampling we will define the superpopulation distribution function by the limit:[2]

$$F(x) \equiv \lim_{N \to \infty} \frac{1}{N} \sum_{i=1}^{N} \mathbb{I}(X_i \leq x) \quad \text{for } x \in \mathbb{R}.$$

We also define the superpopulation mean and variance parameters by:

$$\mu \equiv \int_\mathbb{R} x \, dF(x) \quad \quad \sigma^2 \equiv \int_\mathbb{R} (x - \mu)^2 \, dF(x).$$

We assume that the superpopulation is an exchangeable sequence, allowing us to invoke the representation theorem of de Finetti (see e.g., O'Neill 2009) to treat the population values as IID values from the superpopulation distribution $F$ —i.e., under the model-based approach we have $X_1, X_2, X_3, \ldots \sim$ IID $F$. Exchangeability of the superpopulation implies that the population values are exchangeable, which implies that the sample is a simple-random-sample without replacement from the population. Our analysis in this paper focuses on this common case and is not interested in broader sampling theory for other types of sampling.

---

[1] Note here that we allow for an infinite population by allowing the value $N = \infty$ in the sampling analysis (i.e., we take $N \in \overline{\mathbb{N}}$ as an element of the extented natural numbers).

[2] This expression is the Cesàro limit of the indicators in the sum. There are pathological cases where this Cesàro limit does not exist, but these occur with probability zero under the assumption that $\boldsymbol{X}$ is exchangeable. In those cases we can define $F(x)$ to be the Banach limit for the sequence of indicator values shown in the sum. The latter limit always exists and corresponds to the former when it exists. This definition is discussed in O'Neill (2009, p. 249) in the context of discussion of exchangeable sequences of random variables.

Our analysis will examine the moments of the sample and population, and we will want to be able to refer to the "variance" values for the sample or population with and without Bessel's correction. To do this we will use the following notation:

$$\bar{X}_n \equiv \frac{1}{n}\sum_{i=1}^{n} X_i \qquad \bar{X}_N \equiv \frac{1}{n}\sum_{i=1}^{n} X_i,$$

$$S_n^2 \equiv \frac{1}{n-1}\sum_{i=1}^{n}(X_i - \bar{X}_n)^2 \qquad S_N^2 \equiv \frac{1}{N-1}\sum_{i=1}^{N}(X_i - \bar{X}_N)^2,$$

$$R_n^2 \equiv \frac{1}{n}\sum_{i=1}^{n}(X_i - \bar{X}_n)^2 \qquad R_N^2 \equiv \frac{1}{N}\sum_{i=1}^{N}(X_i - \bar{X}_N)^2.$$

To facilitate later analysis, we also define the (somewhat strange looking) variance quantity:

$$R_{n,N}^2 \equiv \frac{N-1}{N} \cdot S_n^2 = \frac{N-1}{N(n-1)}\sum_{i=1}^{n}(X_i - \bar{X})^2.$$

We will also use the empirical distribution function for the population, which is:

$$F_N(x) \equiv \frac{1}{N}\sum_{i=1}^{N}\mathbb{I}(X_i \leq x) \qquad \text{for } x \in \mathbb{R}.$$

The above notation allows us to differentiate between "the variance" of the population with and without application of Bessel's correction. The quantity $S_N^2$ is the population variance when we apply the correction and the quantity $R_N^2$ is the population variance when we eschew this correction. Different authors have taken different approaches to the proper definition of "the variance" of the population, and we will discuss these approaches soon.

Before proceeding to our analysis of the differences in the two paradigms, it is worth remarking on how our notation differs from the notation in some texts on sampling theory. In applications of the design-based paradigm, it is common to take the quantity $R_N^2$ to define the "population variance", but in that context this quantity is usually labelled as $\sigma^2$, and the population mean is usually labelled as $\mu$ instead of $\bar{X}_N$. (Since there is no superpopulation in the design-based paradigm, the parameters of the superpopulation are not used, avoiding a conflict in notation.) Obviously, the quantity $R_N^2$ is different to the superpopulation parameter $\sigma^2$ (not considered in the design-based approach) so it is important not to equivocate between them. Our notation allows us to refer to all quantities in both paradigms, albeit with different notation than what is used in the design-based paradigm; the reader should note this difference to avoid confusion.

There is substantial variation in notation and nomenclature in textbooks on sampling theory (see e.g., Murthy 1967, Raj 1968, Konijn 1973, Barnett 1974, Cochran 1977, Oñate and Bader 1990, Thompson 2002, Knottnerus 2003). In Table 1 below we present some variation in the names and notations used in various texts, focusing on the notation and name for the population variance. Some authors directly define a quantity that assert to be the "population variance" while others define relevant quantities for their analysis and merely note that others refer to them using terms of this kind.[3] Broadly speaking, authors using the design-based paradigm have tended to use our $R_N^2$ as the "population variance" and sometimes call our $S_N^2$ the "adjusted population variance".

| Reference | Parameter | | |
|---|---|---|---|
| | Name[4] | Notation | Equal to … |
| Murthy (1967), p. 26 | Population variance | $\sigma^2$ | $R_N^2$ |
| Konijn (1973), pp. 2, 34-35 | Variance of values in the population | $\sigma^2$ | $R_N^2$ |
| | Adjusted variance of values in the population | $S^2$ | $S_N^2$ |
| Barnett (1974), p. 24 | Variance of a finite population* | $S^2$ | $S_N^2$ |
| Cochran (1977), p. 23 | The variance … in a finite population is usually defined as… | $\sigma^2$ | $R_N^2$ |
| | …results are presented in terms of a slightly different expression… | $S^2$ | $S_N^2$ |
| Raj (1968), pp. 35-36 (see esp. Remark on p. 36) | Variance of [sampling unit] in the population | $\sigma^2$ | $R_N^2$ |
| | [Not named] | $S_y^2$ | $S_N^2$ |
| Oñate and Bader (1990), pp. 12, 35 | Population variance | $\sigma^2$ | $R_N^2$ |
| Thompson (2002), p. 13 | Finite population variance | $\sigma^2$ | $S_N^2$ |
| Knottnerus (2002), p. 8 | Population variance | $\sigma_y^2$ | $R_N^2$ |

**TABLE 1:** Variation in name/notation for population variance in sampling theory textbooks

---

[3] Although some authors use these mathematical objects without giving them a name (possibly in order to avoid being drawn into the relevant semantic debate), we believe that the absence of explicit naming does not really avoid the issue. Even in the absence of a definite assertion that the relevant measure of spread is a "population variance", a reader will tend to call it *something* (even if only for comprehension of the text) and so the implicit substitution of this name generally obtains in such cases.

[4] Some paraphrasing of names from a broader description has occurred here; the reader should see the relevant references for full details. In particular, Konijn (1973) refers to the variance quantity merely as a "measure of spread" (p. 2) but notes that it is "often also referred to as the variance…" (p. 35). Cochran (1977) prefers to avoid using the term "population variance" himself, but he notes that it is often referred to using this language by others.

## 2. The model-based and design-based paradigms — an illustration

Within our unifying framework, the essential difference between the model-based and design-based paradigms is that the former proceeds without conditioning on the population values, while and the latter does condition on the population values, which is equivalent to conditioning on the empirical distribution $F_N$. This empirical distribution is equivalent to the ordered vector of values in the population. Consequently, conditioning on the empirical distribution of the population means that the analysis in the design-based approach is conditional on the *values* in the population, without knowledge of their *order*. Despite this difference, we will show that both paradigms can be used to derive point estimators and confidence intervals via similar methods, leading to two alternative derivations of the same estimation formulae.

Here we will look at various useful moment results for the two paradigms, and the confidence interval formula for the population mean, which can be derived in two different ways from moment results in the two paradigms. The results in the two paradigms rely on the marginal and conditional moments of the values in the population. Relevant marginal and conditional moments are shown in the following theorems (proved in the appendix). (Here we frame the results in terms of the population quantities $\bar{X}_N$ and $S_N^2$, but they can be framed in terms of $\bar{X}_N$ and $R_N^2$ instead if needed.) It is worth noting that the conditional moment results are stronger than the marginal moment results, and the marginal results can be derived from the conditional moment results using relevant iterated moment formulae (proved in appendix).

**THEOREM 1:** For all $1 \leq i, j \leq N$ we have:

$$\mathbb{E}(X_i) = \mu \qquad \mathbb{V}(X_i) = \sigma^2 \qquad \mathbb{C}(X_i, X_j) = (1 - \mathbb{I}(i \neq j))\sigma^2,$$

$$\mathbb{E}(X_i | F_N) = \bar{X}_N \qquad \mathbb{V}(X_i | F_N) = \frac{N-1}{N} \cdot S_N^2 \qquad \mathbb{C}(X_i, X_j | F_N) = \left(\frac{N-1}{N} - \mathbb{I}(i \neq j)\right) S_N^2.$$

**THEOREM 2:** For all $1 \leq n \leq N$ we have:

$$\mathbb{E}(\bar{X}_n) = \mu \qquad \mathbb{V}(\bar{X}_n) = \frac{\sigma^2}{n} \qquad \mathbb{C}(\bar{X}_n, \bar{X}_N) = \frac{\sigma^2}{N},$$

$$\mathbb{E}(\bar{X}_n | F_N) = \bar{X}_N \qquad \mathbb{V}(\bar{X}_n | F_N) = \frac{N-n}{N} \cdot \frac{S_N^2}{n} \qquad \mathbb{C}(\bar{X}_n, \bar{X}_N | F_N) = 0.$$

**THEOREM 3:** For all $1 \leq n \leq N$ we have:

$$\mathbb{E}(\bar{X}_n - \bar{X}_N) = 0 \qquad \mathbb{V}(\bar{X}_n - \bar{X}_N) = \frac{N-n}{N} \cdot \frac{\sigma^2}{n} \qquad \mathbb{E}(S_n^2) = \sigma^2,$$

$$\mathbb{E}(\bar{X}_n - \bar{X}_N | F_N) = 0 \qquad \mathbb{V}(\bar{X}_n - \bar{X}_N | F_N) = \frac{N-n}{N} \cdot \frac{S_N^2}{n} \qquad \mathbb{E}(S_n^2 | F_N) = S_N^2.$$

**THEOREM 4:** For all $1 \leq n \leq N$ we have:

$$\mathbb{E}(R_n^2) = \frac{n-1}{n} \cdot \sigma^2 \qquad \mathbb{E}(R_n^2 | F_N) = \frac{n-1}{n} \cdot S_N^2,$$

$$\mathbb{E}(R_{n,N}^2) = \frac{N-1}{N} \cdot \sigma^2 \qquad \mathbb{E}(R_{n,N}^2 | F_N) = \frac{N-1}{N} \cdot S_N^2.$$

The above moment results have been framed in terms of the finite population quantities $\bar{X}_N$ and $S_N^2$, but they can of course also be framed in various other ways, depending on the population quantity of interest. To do this, it is useful to introduce some alternative versions of the "finite population correction" term, which we define as:

$$FPC = \frac{N-n}{N} \qquad FPC_* = \frac{N-n}{N-1} \qquad FPC_{**} = \frac{N-n}{N} \cdot \frac{n}{n-1}.$$

The first two of these are quantities that appear sometimes in standard expositions of sampling theory; the third is an alternative that is valuable for presenting an alternative variance quantity. Using these finite population correct terms, it can be shown that the sample quantities defined above satisfy:

$$FPC \cdot S_n^2 = FPC_* \cdot R_{n,N}^2 = FPC_{**} \cdot R_n^2,$$

and the population quantities satisfy:

$$FPC \cdot S_N^2 = FPC_* \cdot R_N^2.$$

Using these finite population correct terms, we can frame our conditional moment results either in terms of $R_N^2$ or $S_N^2$ as shown in the table below. The result of most interest here is the variance of the difference between the sample mean and the population mean. Observe here that both ways of framing this conditional variance involve a scaled "population variance" quantity multiplied by a "finite population correction" term. In the first version (shown on the left) the quantity $R_N^2$ is taken as the "population variance" and the finite population correction is $FPC_*$, whereas in the second version (shown on the right) the quantity $S_N^2$ is taken as the "population variance" and the finite population correction is $FPC$.

| Written in terms of $R_N^2$ | Written in terms of $S_N^2$ |
|---|---|
| $\mathbb{V}(X_i\|F_N) = R_N^2$ | $\mathbb{V}(X_i\|F_N) = \frac{N-1}{N} \cdot S_N^2$ |
| $\mathbb{E}(S_n^2\|F_N) = \frac{N}{N-1} \cdot R_N^2$ | $\mathbb{E}(S_n^2\|F_N) = S_N^2$ |
| $\mathbb{E}(R_n^2\|F_N) = \frac{N(n-1)}{n(N-1)} \cdot R_N^2$ | $\mathbb{E}(R_n^2\|F_N) = \frac{n-1}{n} \cdot S_N^2$ |
| $\mathbb{E}(R_{n,N}^2\|F_N) = R_N^2$ | $\mathbb{E}(R_{n,N}^2\|F_N) = \frac{N-1}{N} \cdot S_N^2$ |
| $\mathbb{V}(\bar{X}_n\|F_N) = FPC_* \cdot \frac{R_N^2}{n}$ | $\mathbb{V}(\bar{X}_n\|F_N) = FPC \cdot \frac{S_N^2}{n}$ |
| $\mathbb{V}(\bar{X}_n - \bar{X}_N\|F_N) = FPC_* \cdot \frac{R_N^2}{n}$ | $\mathbb{V}(\bar{X}_n - \bar{X}_N\|F_N) = FPC \cdot \frac{S_N^2}{n}$ |

Now that we have seen the marginal and conditional moments of interest, we can examine the formulation of a confidence interval for the population mean $\bar{X}_N$. In both paradigms, we form a confidence interval for this quantity via an appropriate pivotal quantity. In the design-based approach we condition on the empirical distribution of the population. In this paradigm, we use the conditional moments of $\bar{X}_n - \bar{X}_N$ and we estimate the population variance $S_N^2$ with the unbiased estimator $S_n^2$ to give $\widehat{\mathbb{V}}(\bar{X}_n - \bar{X}_N|F_N) = FPC \cdot S_n^2/n$. This gives the pivotal quantity:

$$\frac{\bar{X}_n - \bar{X}_N}{\sqrt{\widehat{\mathbb{V}}(\bar{X}_n - \bar{X}_N|F_N)}} = \frac{\sqrt{n}}{\sqrt{FPC}} \cdot \frac{\bar{X}_n - \bar{X}_N}{S_n} \sim \text{St}(n-1).$$

Alternatively, in the model-based approach we proceed without conditioning on the empirical distribution of the population. In this paradigm, we use the marginal moments but we estimate the unknown parameter $\sigma^2$ with the unbiased estimator $S_n^2$ to give $\widehat{\mathbb{V}}(\bar{X}_n - \bar{X}_N) = FPC \cdot S_n^2/n$. This gives the same pivotal quantity:

$$\frac{\bar{X}_n - \bar{X}_N}{\sqrt{\widehat{\mathbb{V}}(\bar{X}_n - \bar{X}_N)}} = \frac{\sqrt{n}}{\sqrt{FPC}} \cdot \frac{\bar{X}_n - \bar{X}_N}{S_n} \sim \text{St}(n-1).$$

In either case, using standard "inversion" techniques (see e.g., O'Neill 2014, p. 286) we can use this pivotal quantity to form a confidence interval for the population mean:

$$\text{CI}_N(1-\alpha) = \left[\bar{X}_n \pm \sqrt{FPC} \cdot \frac{t_{n-1,\alpha/2}}{\sqrt{n}} \cdot s_n\right].$$

Using the fact that $FPC \cdot s_n^2 = FPC_* \cdot r_{n,N}^2 = FPC_{**} \cdot r_n^2$ we can write this confidence interval in either of two alternative forms:

$$\text{CI}_N(1-\alpha) = \left[\bar{X}_n \pm \sqrt{FPC_*} \cdot \frac{t_{n-1,\alpha/2}}{\sqrt{n}} \cdot r_{n,N}\right] = \left[\bar{X}_n \pm \sqrt{FPC_{**}} \cdot \frac{t_{n-1,\alpha/2}}{\sqrt{n}} \cdot r_n\right].$$

Again, we see that the result can be framed in various equivalent ways, either using the variance quantity $S_n^2$, the variance quantity $R_{n,N}^2$ or the variance quantity $R_n^2$ as the appropriate "sample variance" in the confidence interval formula, with corresponding differences in the finite population correction terms. We now proceed to our argument as to which of our quantities has a better claim to being considered as "the population variance".

## 3. What is the appropriate quantity for "the population variance"

In the above exposition we have looked at three general variance quantities $S_n^2$, $R_{n,N}^2$ and $R_n^2$. Substituting $n = N$ we get $S_n^2 = S_N^2$ and $R_{n,N}^2 = R_n^2 = R_N^2$ which reduces things to a choice of two possibilities for "the population variance" — i.e., whether or not to use Bessel's correction. In this paper we will argue that it is more sensible to consider $S_N^2$ to be "the population variance" —i.e., that **we should apply Bessel's correction** when forming the population variance. There are a number of sensible statistical reasons for this view, but before presenting this argument, we will first present the standard argument for the contrary position.

The argument for using $R_N^2$ as "the population variance" is that it is a natural application of the variance operator for a distribution applied to the empirical distribution of the population. To see this, suppose we let $K_i(x) \equiv \mathbb{I}(X_i \leq x)$ denote the indicator functions that appear in our definition of the empirical distribution of the population. This allows us to write the empirical distribution for the population, and its corresponding differential as:

$$F_N(x) = \frac{1}{N}\sum_{i=1}^{N} K_i(x) \quad dF_N(x) = \frac{1}{N}\sum_{i=1}^{N} dK_i(x).$$

The function $K_i$ is the Dirac distribution at the point $X_i$ so we have $\int_\mathbb{R} f(x)\, dK_i(x) = f(X_i)$ for any measurable function $f$. Some simple algebra then gives:

$$\mathbb{E}_{F_N}(f(X)) = \int_\mathbb{R} f(x)\, dF_N(x) = \frac{1}{N}\sum_{i=1}^{N}\int_\mathbb{R} f(x)dK_i(x) = \frac{1}{N}\sum_{i=1}^{N} f(X_i).$$

In particular, we have the following equations:

$$\bar{X}_N = \int_{\mathbb{R}} x \, dF_N(x) \quad \square \quad R_N^2 = \int_{\mathbb{R}} (x - \bar{X}_N)^2 dF_N(x).$$

From these equations we can see that $\bar{X}_N$ and $R_N^2$ are the mean and variance computed relative to the empirical distribution of the population. We have also established that $\mathbb{V}(X_i|F_N) = R_N^2$, which is a natural concomitant of the fact that when we condition on the empirical distribution of the population, the variance of an individual value in the population *is the variance of the empirical distribution*. On this basis, some consider it reasonable to transition from saying that $R_N^2$ is "the variance of the empirical distribution of the population" or "the variance of any population value conditional on the empirical distribution of the population" to saying that $R_N^2$ is the "population variance".

Before proceeding to our own argument to the contrary, we note an immediate non-sequitur in this position. Namely, there is no particular reason that "the variance" of a set of values should be taken to be equivalent to the variance *of the empirical distribution* of those values. That this is so is evident in the inconsistent treatment of the "sample variance", which is universally accepted to incorporate Bessel's correction (which means that it is not equal to the variance of the empirical distribution of the sample). The design-based paradigm gives us an immediate inconsistency in treatment, insofar as it defines "population variance" and "sample variance" by fundamentally different criteria. In the first case "the variance" is taken to be determined by the variance of the empirical distribution of the set of values, but in the second case "the variance" it is determined by the criterion of unbiased estimation of an underlying parameter for an *infinite* population. Of course, one might consider that it is reasonable for a disparity in treatment to exist, because the sample is used precisely *to estimate* a population. One might claim that the sample variance should be defined by its properties as an estimator, precisely because the sample is used for estimation purposes. However, even adopting this view, the design-based paradigm should set the sample variance so that it is an unbiased estimator of the *finite* population variance, which would require the sample variance to be $R_{n,N}^2$, not $S_n^2$. That this position is not adopted shows that this approach does not consistently apply *any* criterion for the meaning of "the variance" of a set of values.

Now, there are three advantages to considering $S_N^2$ to be "the population variance". The first advantage is that —just as with the sample variance— this quantity is an unbiased estimator of the parameter $\sigma^2$, which is the superpopulation variance. This approach means that both the sample variance and population variance are defined by estimation requirements pertaining to parametric estimation in the case of a large population, and the marginal expectation of both quantities is the same. In particular, if we generate a population from an IID model with an underlying fixed distribution (as is often done in simulation work), both the sample variance and population variance will be unbiased estimators of the true variance of the distribution that generated the population. The second advantage is that use of $S_N^2$ as the population variance leads to important moment results and confidence intervals that are framed in terms of the finite-population correction term $FPC$ rather than $FPC_*$, and the former has a much simpler and more natural meaning. The correction term $FPC$ is just the *unsampled proportion of the population*, whereas $FPC_*$ is a transformation of this quantity where Bessel's correction is "shoe-horned" into the correction term. By eschewing Bessel's correction in the population variance, it ends up showing up in the finite population correction term, where it is ill-placed. Finally, the third advantage is that considering $S_N^2$ to be the population variance gives a simple correspondence between the design-based and model-based paradigms, and it is much less likely to lead to confusion for students learning the field.

In the view of the present author, the treatment of $R_N^2$ as "the population variance" is irrational and replete with annoying statistical glitches. The argument in favour of this approach (based on selective appeal to the variance of the empirical distribution) is shallow and unconvincing. Bessel's correction is designed to remove bias from a quantity when it is used as an estimator for a higher-level parameter, but in this approach it is instead built into the correction term for a finite population.

**4. Methodological approaches to defining "the variance" of a set of numbers**

Further to the above analysis, it is worth stepping back and asking a broader methodological question — how should be define "the variance" of a set of numbers? The variance operator is well-defined for a *distribution*, but it is not obvious how it should apply to a vector of *values* generated from a distribution. As we have pointed out, the approach of using the variance of the empirical distribution —proportionate to the sum of squared-deviations from the mean of the values— is one possibility, but it is a non sequitur to assume that this is the only coherent method. Here we will consider variants that allow a different scaling proportion, allowing for unbiased estimation of various quantities that might be of interest.

If we restrict attention to quantities that are proportionate to the sum of squared-deviations from the mean of the values, then it remains only to determine an appropriate scaling constant for the sum-of-squares. In Table 2 below we set out three methodological approaches that give a coherent method of defining "the variance", along with the resulting quantities for the sample variance and population variance. Using the variance of the empirical distribution is one of the methods in the table, but we also show two other coherent methods; none lead to the sample variance $S_n^2$ and the population variance $R_N^2$.

| Methodological Approach | Sample variance | Population variance | FPC |
|---|---|---|---|
| Variance of a set of values is the variance of the empirical distribution of those values | $R_n^2$ | $R_N^2$ | $FPC_{**}$ |
| Variance of a set of values is an unbiased estimator of the variance of the empirical distribution of the population | $R_{n,N}^2$ | $R_N^2$ | $FPC_*$ |
| Variance of a set of values is an unbiased estimator of the superpopulation variance | $S_n^2$ | $S_N^2$ | $FPC$ |
| No consistent methodological approach to the variance of a set of values gives this | $S_n^2$ | $R_N^2$ | * |

**TABLE 2:** Methodological approaches to defining "the variance" of a set of values

Of the three methodological choices shown in the table, only the proposed sample variance values $S_n^2$ and $R_{n,N}^2$ have stable conditional expectations; the conditional expected values of these quantities are $\mathbb{E}(S_n^2|F_N) = S_N^2$ and $\mathbb{E}(R_{n,N}^2|F_N) = R_N^2$ respectively. The other position, using the variance of the empirical distribution, does not yield a stable expected value, and this is undesirable for use in estimation work. Of the two positions with stable expectations, the proposed sample variance $R_{n,N}^2$ has an obvious drawback, in that it is not fully determined by the sample — it is also affected by the population size $N$. This is necessary in order for the quantity to preserve a stable conditional expectation at $R_N^2$, but it is a disqualifying property for a quantity that is proposed as a summary of the variance of the sample. Consequently, in the view of the present author, the position that makes the most sense is to regard $S_n^2$ as the sample variance and $S_N^2$ as the population variance. This position has a stable conditional expectation which is in turn an unbiased estimator of the superpopulation variance.

There are some authors who take the position of treating $S_n^2$ as the sample variance and $R_N^2$ as the population variance. As noted, this position has no coherent methodological approach. But most damning of all is that this position entails use of a quantity that has a sudden "jump down" when we increase the sample size up to the population size (i.e., when we take a full census of the population). Under this incoherent position, the variance of a vector $\boldsymbol{X}_n = (X_1, \ldots, X_n)$ is:

$$\text{Variance of } \boldsymbol{X}_n = \begin{cases} S_n^2 & \text{for } n < N, \\ R_n^2 & \text{for } n = N, \end{cases}$$

and its marginal expectation is:

$$\mathbb{E}(\text{Variance of } \boldsymbol{X}_n) = \begin{cases} \sigma^2 & \text{for } n < N, \\ \dfrac{n-1}{n} \cdot \sigma^2 & \text{for } n = N. \end{cases}$$

In both equations, the variance of the set of values (or its expectation) chugs along with the formula $S_n^2$ (with expectation $\sigma^2$), but then it suddenly "jumps down" to $R_n^2$ (with expectation $(n-1)/n \cdot \sigma^2$) for one particular point. As can be seen, this gives a complicated, inconsistent, and *ad hoc* answer to the simple question: what is the variance of a set of numbers? To say that the sudden "jump" in the answer is a drawback of the position is rather an understatement.

Defenders of the view that $R_N^2$ is "the population variance" might retort that the above question is problematic, because it asks about a quantity (the variance of $\boldsymbol{X}_n$) that is mathematically ambiguous. They might therefore claim that this is *not* a simple question, and their strange answer reflects the ambiguity in the meaning of "the variance". But really, shouldn't this be simple? Can statisticians honestly not give a single consistent formula for "the variance" of a set of numbers without a sudden "jump" occurring in the formula at one particular sample size?

**5. Clearing the cobwebs — suggestions for sampling theory notation and practice**

The confusing nature of the design-based paradigm (for people new to the field) is exacerbated by a number of notational conventions that are at odds with other fields of statistics. The first problem is that all analysis in the paradigm is conditional on $F_N$, but the mathematical results shown in texts leave this conditioning *implicit*, thereby giving readers the false impression that the moment results are marginal moments. (In this paper we have adopted the practice of using explicit conditioning on $F_N$ to make the distinction between marginal and conditional moment results clear.) The second problem is that most texts on design-based analysis use $\mu$ and $\sigma^2$ as the notation for the quantities $\bar{X}_N$ and $R_N^2$, which is at odds with the model-based convention of using Greek letters to denote model "parameters" (i.e., limiting quantities pertaining to the infinite superpopulation). This leads to confusion between distinct quantities for students who are new to the field, and in some cases, it also confuses experienced practitioners.

This latter aspect of the design-based paradigm is probably one of the most confusing aspects of the paradigm for students. Students often encounter model-based statistical problems where they have a set of IID random variables from a distribution —e.g., $X_1, \ldots, X_N \sim \text{IID } N(\mu, \sigma^2)$. To then find within the design-based paradigm that the values $\mu$ and $\sigma^2$ refer to something distinct from the model mean and variance invites confusion. This confusion is compounded by defining "the population variance" without Bessel's correction but "the sample variance" with Bessel's correction, and it is still further compounded by an unnatural finite-population correction term that "shoe-horns" the correction back into the inference formulae.

To give an idea of the kind of conceptual mine-field that awaits students in the field, consider the fact that both the model-based paradigm and the design-based paradigm (each using their own notation) assert various moment results that lack explicit conditioning on the population values, and use parametric notation to mean different things. In the model-based paradigm this is because the moment results are actually marginal moments and the parametric notation refers to parameters in the superpopulation. However, in the design-based paradigm this occurs due to a choice use implicit conditioning in all results and to use parametric notation for (finite) population quantities. Typical of the kinds of assertions one encounters are the following.

|  | **Model-based paradigm** | **Design-based paradigm** |
|---|---|---|
| What they assert (in their own notation) | $\mathbb{E}(\bar{X}_n) = \mu$ | $\mathbb{E}(\bar{X}_n) = \mu$ |
| What they mean (in our notation) | $\mathbb{E}(\bar{X}_n) = \mu$ | $\mathbb{E}(\bar{X}_n \mid F_N) = \bar{X}_N$ |
| What they assert (in their own notation) | $\mathbb{E}(S_n^2) = \sigma^2$ | $\mathbb{E}(S_n^2) = \dfrac{N}{N-1} \cdot \sigma^2$ |
| What they mean (in our notation) | $\mathbb{E}(S_n^2) = \sigma^2$ | $\mathbb{E}(S_n^2 \mid F_N) = \dfrac{N}{N-1} \cdot R_N^2$ |

| What they assert (in their own notation) | $\mathbb{V}(\bar{X}_n) = \dfrac{N-n}{N} \cdot \dfrac{\sigma^2}{n}$ | $\mathbb{V}(\bar{X}_n) = \dfrac{N-n}{N-1} \cdot \dfrac{\sigma^2}{n}$ |
|---|---|---|
| What they mean (in our notation) | $\mathbb{V}(\bar{X}_n) = \dfrac{N-n}{N} \cdot \dfrac{\sigma^2}{n}$ | $\mathbb{V}(\bar{X}_n \mid F_N) = \dfrac{N-n}{N-1} \cdot \dfrac{R_N^2}{n}$ |

**TABLE 3:** Various assertions in the model-based and design-based paradigms

It is difficult to overstate the immense confusion this duality in meaning causes for students (and even many practitioners) who are used to seeing random variables generated by models with unobservable model parameters and labelled using Greek letters. Observe in the above table that one sees pairs of inconsistent equations that appear to be identical except for small differences in scaling constants. These apparently similar equations actually mean completely different things and not only are they not inconsistent — they are not even referring to the same moment quantities or the same "parameters"!

This confusion can be avoided —as it has here— with some simple substantive and notational practices that unify the two paradigms and allow them to be presented as a consistent whole. Our suggestions for exposition of both sampling theory paradigms can be summed up in the following simple injunctions:

1. The notation $\mu$ and $\sigma^2$ (and any other Greek letters) should be reserved only for the superpopulation parameters (i.e., for limiting moments for the infinite population);
2. All probability and moment statements in the design-based analysis should explicitly condition on the population empirical distribution $F_N$;
3. The variance of the sample and population in both paradigms should include Bessel's correction (i.e., we should use $S_n^2$ and $S_N^2$ respectively);
4. The analysis should allow the value $N = \infty$ so that there is an explicit special case for dealing with the superpopulation; and
5. Consequently, the population mean $\bar{X}_N$ and population variance $S_N^2$ can be regarded as generalisations of the superpopulation parameters $\mu$ and $\sigma^2$, occurring when $N = \infty$.

The above suggestions lead to a unifying framework for sampling theory that can draw on the results in either the model-based or design-based paradigm, without confusing the two. These suggestions also make it easy to deal explicitly with inferences for a finite population or an infinite superpopulation. To assist sound understanding of the effect of population size, we also recommend that exposition of results in sampling theory be presented with this three-level analysis (sample-population-superpopulation) at its base. Specifically, our suggested notation for the statements made in Table 2 is shown in Table 4 below.

|  | **Model-based paradigm** | **Design-based paradigm** |
|---|---|---|
| Our suggested notation | $\mathbb{E}(\bar{X}_n) = \mu$ | $\mathbb{E}(\bar{X}_n | F_N) = \bar{X}_N$ |
| Our suggested notation | $\mathbb{E}(S_n^2) = \sigma^2$ | $\mathbb{E}(S_n^2 | F_N) = S_N^2$ |
| Our suggested notation | $\mathbb{V}(\bar{X}_n) = \dfrac{N-n}{N} \cdot \dfrac{\sigma^2}{n}$ | $\mathbb{V}(\bar{X}_n | F_N) = \dfrac{N-n}{N} \cdot \dfrac{S_N^2}{n}$ |

**TABLE 4:** Our suggested notation for the model-based and design-based paradigms

## 6. Concluding remarks

In this paper we have set up a unifying notation that can be used to describe both the model-based and design-based paradigms in sampling theory. This is not a revolutionary change — all we have done is to ensure distinct notation for distinct quantities that can appear in the two paradigms, and we have taken care to always condition *explicitly* on the empirical distribution of the population within the design-based paradigm. Our framework allows both the model-based and design-based paradigms to be described accurately, without notational confusion between them or other objects. We have also suggested that it is appropriate to incorporate Bessel's correction into both the sample variance and population variance; this has a consistent methodological basis, simplifies various sampling theory results, and avoids some annoying statistical glitches that occur under alternative treatments. In particular, we find no merit in the common practice of treating the population variance as a quantity that does not incorporate Bessel's correction — this commonly used convention is irrational and engenders a number of statistical glitches in subsequent analysis.

Exposition of specific results in sampling theory (e.g., setting up confidence intervals) usually proceeds according to one or other of the two paradigms considered here. Within the standard framework involving simple random sampling without replacement from a population, both methods lead to the same confidence intervals via different approaches. The design-based paradigm uses conditional results to directly establish quantities involving the finite population values and then turns these into pivotal quantities by estimating population moments with the corresponding sample moments. Contrarily, the model-based paradigm uses marginal results to establish quantities involving the parameter values from the superpopulation and then turns these into pivotal quantities by estimating the superpopulation parameters with corresponding sample moments. Both approaches are intuitively reasonable and use analogous methods.

We hope that the present paper will provide some clarity to students and practitioners when seeking to understand differences in the two main paradigms for sampling theory. Both of these paradigms give useful results that assist in understanding the dynamics of sampling from a population and making inferences about population quantities. However, as a general rule, the field is a horrible mess due to notational conflicts and a silly choice of definition of the population variance. We hope that practitioners will take a unified view of the field and use notation and substantive definitions that make the subject clear and unambiguous.

**References**

BARNETT, V. (1974) *Elements of Sampling Theory*. The English Universities Press Ltd: London.

COCHRAN, W.G. (1977) *Sampling Techniques (Third Edition)*. John Wiley and Sons: New York.

KNOTTNERUS, P. (2003) *Sample Survey Theory: Some Pythagorean Perspectives*. Springer: New York.

KONIJN, H.S. (1973) *Statistical Theory of Sample Survey Design and Analysis*. North-Holland Publishing Company: Amsterdam.

MURTHY, M.N. (1967) *Sampling Theory and Methods*. Statistical Publishing Society: Calcutta.

OÑATE, B.T, AND BADER, J.M.O. (1990) *Sampling Surveys and Applications*. College: Laguna.

O'NEILL, B. (2009) Exchangeability, correlation and Bayes' effect. *International Statistical Review* **77(2)**, pp. 241-250.

O'NEILL, B. (2014) Some useful moment results for sampling problems. *The American Statistician* **68(4)**, pp. 282-296.

RAJ, D. (1968) *Sampling Theory*. McGraw-Hill Book Company: New York.

THOMPSON, S.K. (2002) *Sampling (Second Edition)*. Wiley: New York.

# Appendix: Proofs of moment results

Here we set out proofs of the theorems in the main body of the paper. Our approach here is to prove the conditional moment results and then use these to prove the marginal moment results from these using the laws of iterated moments.

**LEMMA 1:** We have:

$$\sum_{k=1}^{N} X_k^2 = N(\bar{X}_N^2 + R_N^2).$$

**PROOF OF LEMMA 1:** Expanding the quadratic shows that:

$$NR_N^2 = \sum_{k=1}^{N}(X_k - \bar{X}_N)^2 = \sum_{k=1}^{N} X_k^2 - 2\sum_{k=1}^{N} X_k \bar{X}_N + \sum_{k=1}^{N} \bar{X}_N^2$$

$$= \sum_{k=1}^{N} X_k^2 - 2N\bar{X}_N^2 + N\bar{X}_N^2$$

$$= \sum_{k=1}^{N} X_k^2 - N\bar{X}_N^2.$$

Re-arranging this equation gives the result in the lemma. ∎

**PROOF OF THEOREM 1:** Since the sequence $X$ is exchangeable, this means that $X_i$ is a random value from the population. Hence, taking $I \sim \text{U}\{1, \ldots, N\}$ to be a random index for a value in the population, we have conditional mean:

$$\mathbb{E}(X_i|F_N) = \mathbb{E}(X_I|F_N) = \sum_{k=1}^{N} X_k \cdot \mathbb{P}(I=k) = \frac{1}{N}\sum_{k=1}^{N} X_k = \bar{X}_N,$$

and conditional variance:

$$\mathbb{V}(X_i|F_N) = \mathbb{V}(X_I|F_N) = \sum_{k=1}^{N}(X_k - \bar{X}_N)^2 \cdot \mathbb{P}(I=k) = \frac{1}{N}\sum_{k=1}^{N}(X_k - \bar{X}_N)^2 = \frac{N-1}{N} \cdot S_N^2.$$

This establishes the first two moment results, so it remains to prove the third result. For the case where $i = j$ we have $\mathbb{C}(X_i, X_j|F_N) = \mathbb{C}(X_i, X_i|F_N) = \mathbb{V}(X_i|F_N) = R_N^2$, which matches the formula in the theorem. To deal with the case where $i \neq j$, take $J \sim \text{U}\{1, \ldots, i-1, i+1, \ldots N\}$ to be a random index for a value in the population —distinct from the first index— we have:

$$\mathbb{E}(X_i X_j | F_N) = \mathbb{E}(X_I X_J | F_N) = \sum_{k=1}^{N} \sum_{\ell=1}^{N} X_k X_\ell \cdot \mathbb{P}(I = k, J = \ell)$$

$$= \sum_{k=1}^{N} \sum_{\ell=1}^{N} X_k X_\ell \cdot \frac{1}{N} \cdot \frac{1}{N-1} \cdot \mathbb{I}(I \neq J)$$

$$= \frac{1}{N(N-1)} \sum_{k=1}^{N} \sum_{\ell \neq k}^{} X_k X_\ell$$

$$= \frac{1}{N(N-1)} \left[ \sum_{k=1}^{N} \sum_{\ell=1}^{N} X_k X_\ell - \sum_{k=1}^{N} X_k^2 \right]$$

$$= \frac{1}{N(N-1)} \left[ N^2 \bar{X}_N^2 - \sum_{k=1}^{N} X_k^2 \right]$$

$$= \frac{1}{N(N-1)} [N^2 \bar{X}_N^2 - N(\bar{X}_N^2 + R_N^2)]$$

$$= \frac{1}{N(N-1)} [N(N-1) \bar{X}_N^2 - N R_N^2]$$

$$= \bar{X}_N^2 - \frac{1}{N-1} R_N^2$$

$$= \bar{X}_N^2 - \frac{1}{N} \cdot S_N^2.$$

We then have:

$$\mathbb{C}(X_i, X_j | F_N) = \mathbb{C}(X_I, X_J | F_N) = \mathbb{E}(X_I X_J | F_N) - \mathbb{E}(X_I | F_N)^2$$

$$= \bar{X}_N^2 - \frac{1}{N} \cdot S_N^2 - \bar{X}_N^2$$

$$= -\frac{1}{N} \cdot S_N^2$$

$$= \left( \frac{N-1}{N} - 1 \right) S_N^2,$$

This establishes the conditional moment formulae. It is simple to prove these results directly, but here we will prove them via use of the conditional moment formulae above, to illustrate the connection noted in the paper. Applying the laws of iterated expectation, variance, and covariance (with $F_N$ as the relevant random object), we obtain:

$$\mathbb{E}(X_i) = \mathbb{E}(\mathbb{E}(X_i | F_N))$$
$$= \mathbb{E}(\bar{X}_N) = \mu.$$

$$\mathbb{V}(X_i) = \mathbb{E}(\mathbb{V}(X_i | F_N)) + \mathbb{V}(\mathbb{E}(X_i | F_N))$$
$$= \mathbb{E}(R_N^2) + \mathbb{V}(\bar{X}_N)$$

$$= \frac{N-1}{N} \cdot \sigma^2 + \frac{\sigma^2}{N}$$

$$= \sigma^2 - \frac{\sigma^2}{N} + \frac{\sigma^2}{N} = \sigma^2.$$

$$\mathbb{C}(X_i, X_j) = \mathbb{E}(\mathbb{C}(X_i, X_j | F_N)) + \mathbb{C}(\mathbb{E}(X_i | F_N), \mathbb{E}(X_j | F_N))$$

$$= \left(1 - \frac{N}{N-1} \cdot \mathbb{I}(i \neq j)\right) \mathbb{E}(R_N^2) + \mathbb{C}(\bar{X}_N, \bar{X}_N)$$

$$= \left(1 - \frac{N}{N-1} \cdot \mathbb{I}(i \neq j)\right) \frac{N-1}{N} \cdot \sigma^2 + \mathbb{V}(\bar{X}_N)$$

$$= \left(1 - \frac{N}{N-1} \cdot \mathbb{I}(i \neq j)\right) \frac{N-1}{N} \cdot \sigma^2 + \frac{\sigma^2}{N}$$

$$= \frac{N-1}{N} \cdot \sigma^2 - \mathbb{I}(i \neq j) \cdot \sigma^2 + \frac{\sigma^2}{N}$$

$$= (1 - \mathbb{I}(i \neq j))\sigma^2.$$

This establishes the marginal moment results in the theorem. ∎

**PROOF OF THEOREM 2:** Applying the conditional expectation result in Theorem 1 gives:

$$\mathbb{E}(\bar{X}_n | F_N) = \mathbb{E}\left(\frac{\sum_{k=1}^n X_i}{N} \bigg| F_N\right)$$

$$= \frac{1}{n} \sum_{k=1}^n \mathbb{E}(X_i | F_N)$$

$$= \frac{1}{n} \sum_{k=1}^n \bar{X}_N = \bar{X}_N.$$

Applying the conditional covariance result in Theorem 1 gives:

$$\mathbb{V}(\bar{X}_n | F_N) = \mathbb{V}\left(\frac{\sum_{i=1}^n X_i}{n} \bigg| F_N\right)$$

$$= \frac{1}{n^2} \sum_{i=1}^n \sum_{j=1}^n \mathbb{C}(X_i, X_j | F_N)$$

$$= \frac{1}{n^2} \sum_{i=1}^n \sum_{j=1}^n \left(1 - \frac{N}{N-1} \cdot \mathbb{I}(i \neq j)\right) R_N^2$$

$$= \frac{1}{n^2} \left[n^2 - \frac{N}{N-1} \cdot n(n-1)\right] R_N^2$$

$$= \left[n - \frac{N}{N-1} \cdot (n-1)\right] \frac{R_N^2}{n}$$

$$= \left[ \frac{n(N-1)}{N-1} - \frac{N(n-1)}{N-1} \right] \frac{R_N^2}{n}$$

$$= \frac{N-n}{N-1} \cdot \frac{R_N^2}{n}$$

$$= \frac{N-n}{N} \cdot \frac{S_N^2}{n}.$$

Applying the conditional covariance result in Theorem 1 gives:

$$\mathbb{C}(\bar{X}_n, \bar{X}_N | F_N) = \mathbb{C}\left( \frac{\sum_{i=1}^n X_i}{n}, \frac{\sum_{i=1}^N X_i}{N} \bigg| F_N \right)$$

$$= \frac{1}{nN} \sum_{i=1}^n \sum_{j=1}^N \mathbb{C}(X_i, X_j | F_N)$$

$$= \frac{1}{nN} \sum_{i=1}^n \sum_{j=1}^N \left( 1 - \frac{N}{N-1} \cdot \mathbb{I}(i \neq j) \right) R_N^2$$

$$= \frac{1}{nN} \left[ nN - \frac{N}{N-1} \cdot n(N-1) \right] R_N^2$$

$$= \frac{1}{nN} [nN - nN] R_N^2 = 0.$$

It is simple to prove these results directly, but here we will prove them via use of the conditional moment formulae above, to illustrate the connection noted in the paper. Applying the laws of iterated expectation, variance, and covariance (with $F_N$ as the random object), we obtain:

$$\mathbb{E}(\bar{X}_n) = \mathbb{E}(\mathbb{E}(\bar{X}_n | F_N))$$

$$= \mathbb{E}(\bar{X}_N) = \mu.$$

$$\mathbb{V}(\bar{X}_n) = \mathbb{E}(\mathbb{V}(\bar{X}_n | F_N)) + \mathbb{V}(\mathbb{E}(\bar{X}_n | F_N))$$

$$= \mathbb{E}\left( \frac{N-n}{N-1} \cdot \frac{R_N^2}{n} \right) + \mathbb{V}(\bar{X}_N)$$

$$= \frac{N-n}{N} \cdot \frac{\sigma^2}{n} + \frac{\sigma^2}{N}$$

$$= \frac{\sigma^2}{n} - \frac{\sigma^2}{N} + \frac{\sigma^2}{N} = \frac{\sigma^2}{n}.$$

$$\mathbb{C}(\bar{X}_n, \bar{X}_N) = \mathbb{E}(\mathbb{C}(\bar{X}_n, \bar{X}_N | F_N)) + \mathbb{C}(\mathbb{E}(\bar{X}_n | F_N), \mathbb{E}(\bar{X}_N | F_N))$$

$$= \mathbb{E}(0) + \mathbb{C}(\bar{X}_N, \bar{X}_N)$$

$$= \mathbb{V}(\bar{X}_N)$$

$$= \frac{\sigma^2}{N}.$$

This establishes the marginal moment results in the theorem. ∎

**PROOF OF THEOREM 3:** Applying the results in Theorem 2 we have:

$$\mathbb{E}(\bar{X}_n - \bar{X}_N | F_N) = \mathbb{E}(\bar{X}_n | F_N) - \mathbb{E}(\bar{X}_N | F_N)$$
$$= \mu - \mu = 0.$$

$$\mathbb{V}(\bar{X}_n - \bar{X}_N | F_N) = \mathbb{V}(\bar{X}_n | F_N) - 2\mathbb{C}(\bar{X}_n, \bar{X}_N | F_N) + \mathbb{V}(\bar{X}_N | F_N)$$
$$= \frac{N-n}{N-1} \cdot \frac{R_N^2}{n} - 0 + 0$$
$$= \frac{N-n}{N-1} \cdot \frac{R_N^2}{n}$$
$$= \frac{N-n}{N} \cdot \frac{S_N^2}{n}.$$

To prove the last conditional moment result, we first note that:

$$\sum_{i=1}^{n} (X_i - \bar{X}_n)^2 = \sum_{i=1}^{n} X_i^2 - 2\bar{X}_n \sum_{i=1}^{n} X_i + n\bar{X}_n^2 = \sum_{i=1}^{n} X_i^2 - n\bar{X}_n^2.$$

We also have:

$$\mathbb{E}(X_i^2 | F_N) = \mathbb{V}(X_i | F_N) + \mathbb{E}(X_i | F_N)^2 = R_N^2 + \bar{X}_N^2,$$

$$\mathbb{E}(\bar{X}_n^2 | F_N) = \mathbb{V}(\bar{X}_n | F_N) + \mathbb{E}(\bar{X}_n | F_N)^2 = \frac{N-n}{N-1} \cdot \frac{R_N^2}{n} + \bar{X}_N^2.$$

Applying these results, we have:

$$\mathbb{E}(\sum_{i=1}^{n}(X_i - \bar{X}_n)^2 | F_N) = \mathbb{E}(\sum_{i=1}^{n} X_i^2 - n\bar{X}_n^2 | F_N)$$
$$= \sum_{i=1}^{n} \mathbb{E}(X_i^2 | F_N) - n\mathbb{E}(\bar{X}_n^2 | F_N)$$
$$= nR_N^2 + n\bar{X}_N^2 - \frac{N-n}{N-1} \cdot R_N^2 - n\bar{X}_N^2$$
$$= nR_N^2 - \frac{N-n}{N-1} \cdot R_N^2$$
$$= \frac{n(N-1) - (N-n)}{N-1} \cdot R_N^2$$
$$= N \cdot \frac{n-1}{N-1} \cdot R_N^2.$$

We therefore have:

$$\mathbb{E}(S_n^2 | F_N) = \frac{N}{N-1} \cdot R_N^2 = S_N^2.$$

It is simple to prove the remaining results directly, but here we will prove them via use of the conditional moment formulae, so as to illustrate the connection noted in the paper. Applying the laws of iterated expectation, variance, and covariance (with $F_N$ as the relevant random object), we obtain:

$$\mathbb{E}(\bar{X}_n - \bar{X}_N) = \mathbb{E}(\mathbb{E}(\bar{X}_n - \bar{X}_N | F_N))$$
$$= \mathbb{E}(0) = 0,$$

$$\mathbb{V}(\bar{X}_n - \bar{X}_N) = \mathbb{E}(\mathbb{V}(\bar{X}_n - \bar{X}_N | F_N)) + \mathbb{V}(\mathbb{E}(\bar{X}_n - \bar{X}_N | F_N))$$
$$= \mathbb{E}\left(\frac{N-n}{N} \cdot \frac{S_N^2}{n}\right) + \mathbb{V}(0)$$
$$= \frac{N-n}{N} \cdot \frac{S_N^2}{n},$$

$$\mathbb{E}(S_n^2) = \mathbb{E}(\mathbb{E}(S_n^2 | F_N))$$
$$= \mathbb{E}(S_N^2) = \sigma^2.$$

This establishes the moment results in the theorem. ∎

**PROOF OF THEOREM 4:** Since $R_n^2 = (n-1)/n \times S_n^2$ and $R_{n,N}^2 = (N-1)/N \times S_n^2$ these results trivial consequences of the sample variance results in Theorem 3. Using those results we have:

$$\mathbb{E}(R_{n,N}^2) = \mathbb{E}\left(\frac{N-1}{N} \cdot S_n^2\right) = \frac{N-1}{N} \cdot \mathbb{E}(S_n^2) = \frac{N-1}{N} \cdot \sigma^2,$$

$$\mathbb{E}(R_{n,N}^2 | F_N) = \mathbb{E}\left(\frac{N-1}{N} \cdot S_n^2 \Big| F_N\right) = \frac{N-1}{N} \cdot \mathbb{E}(S_n^2 | F_N) = \frac{N-1}{N} \cdot S_N^2.$$

This establishes the moment results in the theorem. ∎

$$\mathbb{E}(R_n^2) = \mathbb{E}\left(\frac{n-1}{n} \cdot S_n^2\right) = \frac{n-1}{n} \cdot \mathbb{E}(S_n^2) = \frac{n-1}{n} \cdot \sigma^2,$$

$$\mathbb{E}(R_n^2 | F_N) = \mathbb{E}\left(\frac{n-1}{n} \cdot S_n^2 \Big| F_N\right) = \frac{n-1}{n} \cdot \mathbb{E}(S_n^2 | F_N) = \frac{n-1}{n} \cdot S_N^2.$$